# PSYCHOLOGICAL ANTECEDENTS TO EMERGENCE OF TEAM AUTONOMY IN AGILE SCRUM TEAMS

Ravikiran Kalluri

Department of Engineering Management and Systems Engineering, Old Dominion University, Norfolk, Virginia, USA

## ABSTRACT

*Agile project management methods are gaining in popularity in the software industry as software development teams are being asked to be adaptive to market needs and resilient to change and uncertainty. With increasing market uncertainty, global competition, and time-to-market pressure, it is becoming a challenge to develop an innovative product and deliver it on-time without the opportunity that comes from team autonomy to experiment and learn from failures. The purpose of this research study was to study the influence of key psychological factors on emergence of Agile team autonomy that leads to Agile project success in software organizations. Using an online survey instrument, the study sampled 137 software professionals from US software companies with experience in the Agile Scrum role of Team Member. The relationship between the human psychology factors pertaining to leadership style, organization structure, human resource practices, customer engagement and Agile team autonomy is explained through multiple linear regression. One-way ANOVA and Pearson's correlation coefficient were used to demonstrate the existence (or nonexistence) of relationships between variables. Finally, an empirical model relating the human psychology factor variables and the dependent variable of Agile team autonomy was constructed for the population.*

## 1. INTRODUCTION

The software engineering domain has become highly complex and the ability to create pathbreaking products needs complex, unbridled human thought. As the work of software engineers becomes more complex, more decisions must be made at the team level than at the leadership level. It is crucial to empower Agile Scrum teams to make independent judgment calls so that software firms can deploy new technologies and effectively manage existing technologies. Taking action and offering Agile Scrum teams the opportunity to grow will evolve their role, strengthen their loyalty towards the organization and promote employee retention.

Empowerment of project teams was shown as an indicator of project development agility by Sheffield and Lemétayer (2013). Hoda et al. (2013), Gill (2014), and Stettina and Horz (2015) have discussed the notion of self-organizing teams, agility of people, processes, tools, and consideration of a revised culture. The authors noted that autonomous teams had increased interaction, were more stable, and experienced increased collaboration, transparency, and trust. Conforto et al. (2014) agree that enablers like flat organizational structure, open culture, and team empowerment are necessary for proper application of agile practices.

According to the self-determination theory by Deci and Ryan (2000), the shift from an external to an internal locus of control encourages more proactive behavior in Agile teams towards achieving shared goals while also making them feel more responsible for the project outcomes. Verwijs and Russo (2023) argue that high-autonomy teams are more inclined to engage in continuous improvement than low-autonomy teams. The increased sense of responsibility for their outcomes will lead high-autonomy teams to show greater concern for the needs of stakeholders and more proactive behaviors aimed at understanding those needs. This will make the high-autonomy teams more responsive than low-autonomy teams due to decreased external dependencies. Specifically for Agile teams, Junker et al. (2022) found that teams are more likely to initiate changes and improvements when team autonomy is high. According to Takeuchi & Nonaka (1986), Scrum teams are more effective and efficient because they are self-organized with overlapping project phases, they learn together as well as transfer learning to the organization, and they receive limited management direction while maintaining subtle control of the project. This method was discussed by Moe et al. (2010) as a new approach for managing projects by providing decision-making authority to the Scrum team members who will be experiencing problems and uncertainties.

According to Malik et al. (2021), Agile team autonomy is a strong antecedent to Agile project success. But there is no single research study that has examined the full spectrum of human psychology factors arising from Leadership Style, Organization Structure, Human Resource Practices and Customer Engagement that influence emergence of Agile team autonomy that ultimately leads to Agile project delivery success. Furthermore, most studies have focused on individual autonomy while team autonomy has been largely neglected (Langfred, 2004; Hodgson & Briand, 2013; Verwijs & Russo, 2023). This research study has attempted to address a significant lacuna in the current body of knowledge on the psychological antecedents of Agile team autonomy. The study results provide insights into the human psychology factors that motivate managerial and team behavior. This can help organizations to design the right incentive programs to promote high work performance and team productivity by human resource departments in software firms.

Following a review of 200 published articles on Agile teams, four prominent human psychology factors emerged - leadership style, organization structure, human resource practices and customer engagement - as the key influencers of Agile team autonomy. These factors are regarded as human psychology factors because they influence team autonomy which is a psychological perception of the team and the individual team members. Exhibit 1 illustrates the variables for this study. The theoretical constructs are summarized in Exhibit 2. These constructs represent the capability and structure of the employees involved with the project as well as the environment in which they complete their project work.

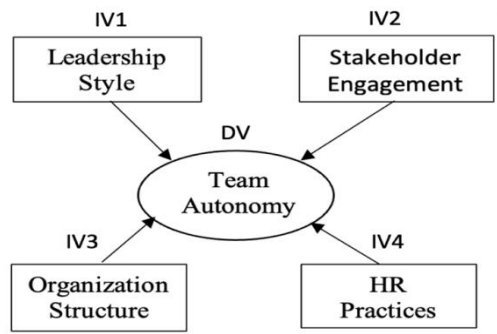

Exhibit 1. Human Psychology Factors That Influence Agile Team Autonomy

Exhibit 2. Construct Summaries

| Name | Definition | References |
| --- | --- | --- |
| Human Psychology Factors | Factors concerned primarily with the psychological effects of leadership style, level of stakeholder (limited to customers for this study) engagement, organization structure, and human resource practices on Agile Scrum team members in a task-oriented environment. | This is a new construct |
| Team Autonomy | A self-directed behavior with general limits set by managerial control, which, if granted, ensures required resources' allocation and encourages employees' trial-and-error experimentations. | (Feldman, 1989) (Zhang et al., 2010) (Verwijs & Russo, 2023) |
| Agile Project Success | Measured by customer satisfaction through predictable delivery of business value. | (Malik et al., 2021) |

## 2. LEADERSHIP STYLE

Research on leadership in agile teams mostly differentiates between a leader as peer or coach to the team who provides appropriate boundary conditions (Takeuchi &Nonaka,1986) and an autonomous team that self-organizes its operational work (Hoda et al., 2013). While some researchers suggest a facilitator who serves as a peer to team members (Takeuchi & Nonaka, 1986) or a leader who empowers the team to lead itself (Manz & Sims, 1987), other researchers do not consider a formal leader of the team but instead emphasize self-organizing roles within the team (Hoda et al., 2013). The successful implementation of self-organized teams into the industrial sector has been of ongoing interest over the last 70 years. Researchers consider the topic from different angles among which are socio-technical systems (Manz & Sims, 1987; Srivastava & Jain, 2017; Trist & Bamforth, 1951), knowledge management (Takeuchi & Nonaka, 1986), complexity theory (Ba̎cklander, 2019; Schwaber, 1997), role theory (Hoda et al., 2013; Yang, 1996) and agile project management (Sutherland & Schwaber, 2017). One recurring topic across the various streams of research is the role of leadership in a team that is by definition self-organized. Elloy (2005) concluded that teams that were led by a supervisor who exhibited the traits of a superleader had higher levels of organization commitment, job satisfaction, and organization self-esteem. Carson et al. (2007) suggest that teams do well when they rely on leadership provided by the team as a whole rather than looking to a single individual to lead them. Werder (2018) finds management support to be associated with self-organization as it can strengthen the forces of self-organization within the team and prevent external forces from limiting self-organization. The level of support is not only a verbal commitment but also the corresponding actions and financial support should follow. Higher extent of self-organization and autonomy helps the team to improve its agility. Flores et al. (2018) showed the moderating role of emotional self-leadership on team decision quality. Through emotional self-leadership, team members can actively anticipate, guide and focus their emotional responses to cognitive conflict and reduce their experience of affective conflict, thus improving team decision quality. Crowder (2015) emphasizes that leadership has to focus on removing roadblocks, encouraging openness and communication, keeping track of the change driven environment to ensure that the overall product meets in goals and requirements, while letting the team take the ground level decisions in the agile development process.

Supportive leadership behaviors at the individual level include leader inclusiveness (Bienefeld &Grote, 2014; Carmeli et al., 2010), support (May et al., 2004), trustworthiness (Madjar & Ortiz-Walters, 2009), openness (Detert & Burris, 2007) and behavioral integrity (Palanski & Vogelgesang, 2011) that strongly influence employee perceptions of team autonomy. At the team level, employees' collective perceptions of support and coaching forwarded by the team leader (Edmondson, 1999; Roberto, 2002), leader inclusiveness (Hirak et al., 2012; Nembhard and Edmondson, 2006), trust in the leader (Li and Tan, 2012; Schaubroeck et al., 2011), and the behavioral integrity of the leader (Leroy et al., 2012) have been found to foster team-level outcomes such as team learning behavior, team performance, engagement in quality improvement work, and reduction in errors among team members. Other work has found that positive leadership styles such as transformational leadership (Nemanich &Vera, 2009), ethical leadership (Walumbwa & Schaubroeck, 2009), change- oriented leadership (Ortega et al., 2014) and shared leadership (Liu et al., 2014) are positively and strongly related to such outcomes as employee voice behavior, team learning, and individual learning. Finally, research has established that leaders who value participation, people, and production use dyadic discovery methods rather than group-based discovery methods (Roussin, 2008; Wong et al., 2010), and an improvement orientation management style (Halbesleben & Rathert, 2008), are able to foster high levels of team autonomy. Based on the literature review, my first research hypothesis is as follows:

- *Hypothesis 1: No significant relationship exists between leadership style and team autonomy.*

## 3. ORGANIZATION STRUCTURE

At the individual level, there is growing evidence that supportive organizational practices are positively related to employee work outcomes such as organizational commitment and job performance. For example, research has found that employee perceptions of organizational support (Carmeli & Zisu, 2009), access to mentoring (Chen et al., 2014), and diversity practices (Singh et al., 2013) foster positive work outcomes. Drawing on a sample of 191 medical professionals in an Israeli medical clinic, supportive organizational practices were found to foster team autonomy through social learning processes, similar to that of supportive leadership behaviors (Carmeli & Zisu, 2009).

Growing research at the individual, team, and organizational levels has looked at social support and the social capital (resources) inherent in relationship networks as key determinants of team self-reliance. At the individual level, research has established that rewarding co-worker relationships and the extent to which members of the organization interact with one another on an interpersonal basis, influence individual learning and engagement (Carmeli &Gittell, 2009; Carmeli et al., 2009; May et al., 2004). Similarly, at the team level, researchers have found that relationship networks, and the social support and resources inherent in such networks, promote team autonomy and contribute to team learning, performance, and innovation. Finally, at the organizational level Carmeli (2007) found that the strength of social networks between teams was positively related to their ability to learn from failure when team autonomy is in place.

At the team level, researchers have found that characteristics such as shared team rewards (Chen & Tjosvold, 2012), formal team structures (Bresman & Zellmer-Bruhn, 2013; Bunderson & Boumgarden, 2010), and team engagement in boundary work (buffering, spanning, and reinforcement) (Faraj & Yan, 2009) are positively associated with higher levels of team autonomy. However, Chandrasekaran & Mishra (2012) found that only team autonomy influenced psychological safety when the project goals and processes of the team were aligned with their broader organizational goals and when there were low degrees of relative exploration (i.e., the team focused on refining existing products and processes rather than seeking to develop new products and processes). Contrary to what they expected, Lau & Murnighan (2005) found that the presence of strong faultlines within teams (i.e., the existence of sub-groups with non-overlapping demographic characteristics) led to greater team autonomy among team members. They argued that this may have resulted from generalization of the positive social effects within strong faultline groups to the entire team. Finally, O'Neill (2009) found that when team members were collectively responsible for bad investment decisions, the presence of team autonomy gave them the courage to admit failure, as compared to when they were individually responsible. The professionally derived status of the team (Nembhard & Edmondson, 2006) leads to outcomes such as the willingness of individuals to speak up and team engagement through enhancing team autonomy. This work suggests that the higher the team autonomy, the safer individuals will feel to experiment and speak up about their innovative ideas.

Malik et al. (2021) collected data to find support for the hypothesized relationships between agile practices, psychological empowerment, innovative behavior, and project performance. The statistical results showed that the agile practices of team autonomy and agile communication contributed to psychological empowerment that led to the innovative behavior of agile teams. The resulting innovative behavior had a significant effect on project performance. Teams themselves can influence the internal organization of teams, but team performance depends not only on the competence of the team itself in managing and executing its work; it also depends on the organizational context provided by management (Stray et al., 2011). Organizational culture - procedures, hierarchical bureaucracy, and traditional mind-set can hinder the performance of agile teams. Digitization is forcing IT organizations to become less hierarchical and more team-based networks. An agile organization has to promote ambidexterity (Lindskog & Magnusson, 2021) practices and support a balanced hybrid structure (Zasa et al., 2020). According to Mateos-Garcia et al. (2008), established 'rational' methodologies in project management emphasize idealized top-down processes that neglect 'soft' human dimensions of projects. This has led to negative outcomes such as delays in delivery, low quality products and overshot budgets.

Based on the literature review, my second research hypothesis is as follows:

- *Hypothesis 2: No significant relationship exists between customer engagement and team autonomy.*

## 4. HUMAN RESOURCE PRACTICES

Zavyalova et al. (2018) demonstrated that agile firms tend to more strongly rely on HRM practices (especially, motivation- and opportunity-enhancing) in ensuring high organizational performance. Furthermore, successful agile firms had more centralized HRM architectures with less authority diffusion among different levels of management. The study shows that with the elimination of line managers in agile organizations, HR functions are being transferred back to HRM departments. HR is instrumental in establishing the ethics and compliance policies that among other things ensure psychological safety (Edmondson, 1999) and autonomy for Agile Scrum teams. The strong correlation between the Agile project management approach and HRM architecture may be a crucial reason for why many agile transformations in project-based organizations have failed. The psychological ramifications of Agile transformation in a large enterprise cannot be neglected. Grass et al. (2020) identified empowerment as a focal human factor for adaptability emergence in teams. The findings demonstrate that empowerment is not a static state, but rather emerges through the interactions between various actors. Specifically, the team and its leader engage in both empowerment-enhancing and empowerment-reducing activities. These activities are further influenced by the agile team's immediate context: Two-fold customer influences, that is, supporting and hindering empowerment interactions, the organizational environment, that is, undergoing an agile transformation and supportive top management behaviors.

Owusu (1999) emphasizes the importance of employee involvement and human resource development in agile management systems coupled with good communication. According to Muduli (2016), organizations have to design practices related to organizational learning and training, compensation, involvement, teamwork and IS and implement them efficiently and effectively to enable agility within the workforce, as only an agile workforce can respond proactively to a volatile, uncertain, complex, and ambiguous business environment. Further, the study also suggests that managers should design the organizational practices capable of enhancing psychological empowerment, as the combination can deliver better workforce agility. Holbeche (2018) argues that HR can help embed organizational agility by creating holistic scorecards of the right performance metrics that link to the vision. Performance management systems and other people practices should reinforce the Agile principles and become developmental in orientation. For collaboration (with real accountability), promotion criteria should require evidence of collaboration and effective delivery – with job rotations required for moving up the ladder. Similarly, recognition and reward mechanisms should reinforce desirable cultural practices that are central to developing an agile culture. Hennel & Rosenkranz (2021) conclude that social agile practices promoted by HR positively influence autonomy, transparency, communication, and ultimately productivity in Agile teams. Based on the literature review, my third research hypothesis is as follows:

- *Hypothesis 3: No significant relationship exists between organization structure and team autonomy.*

## 5. CUSTOMER ENGAGEMENT

Power (2010) shows how agile practices such as sprint demos facilitate early and continuous involvement of customers in the development process and provide an opportunity to address their needs. The team needs to be empowered to make rapid course corrections in light of any new customer feedback. Leadership involvement can slow down the responsiveness of the team that has the best context of the problem and the best chance of finding an innovative solution. Denning (2017) says the right question starts with the customer. To meet customers' needs through enterprise agility, the thinking part of an organization must be run as a network, in which information flows horizontally and upwards, as well as downwards. Work in a network is mostly done in self-organizing teams that are in pursuit of the firm's overall goal. Based on the literature review, my fourth research hypothesis is as follows:

- *Hypothesis 4: No significant relationship exists between human resource practices and team autonomy.*

### METHOD

*Sample*

The study employed a convenience sample of 137 software engineers from US software companies with experience in the Agile Scrum role of Team Member. The convenience sample was however sufficiently randomized by the anonymous nature of the survey data collection. The target sample size for the study was 300 but the survey was able to get sufficient data from 137 responses (Bordens & Abbott, 2011). The primary target for outreach were professional associations like The American Society of Engineering Management and Scrum.org. Membership includes software practitioners that hold a variety of positions within their companies. This ensured internal validity of the data being collected from a broad spectrum of US software industry professionals. These software professionals can work in large corporations as well as start-ups in the software industry. An auxiliary medium for outreach was via social media i.e., Linked-in groups like IEEE Software and SCRUM study associated with Agile project management body of knowledge and my professional network. All participants had at least one year of Agile Scrum work experience. It was made clear in the survey invite letter that participation is purely voluntary. The survey input data did not require participants to disclose any personal or confidential information.

**MEASURES**

*Leadership*

The study used the Leader Empowering Behavior Questionnaire (LEBQ) developed by Konczak et al. (2000). LEBQ consists of 17 items under six proposed dimensions of leader-empowering behaviors. The LEBQ contains 17 items grouped in six dimensions (three items per construct, except for one of them):

(1) Delegation of authority refers to whether the leaders grant power to subordinates.
(2) Accountability for outcomes addresses the leader's emphasis on taking responsibility for consequences.
(3) Self-directed decision making implies that the leader encourages independent decision-making.
(4) Information sharing evaluates whether the leaders share information and knowledge with the employees.
(5) Skill development is concerned with the extent to which the leader facilitates the development of skills and secures appropriate training for employees.
(6) Coaching for innovative performance is related to behavior that encourages calculated risk-taking and new ideas and provides performance feedback to employees, treating their mistakes and setbacks as opportunities to learn.

The LEBQ is answered on a Likert-type scale that ranges from 1 ("strongly disagree") to 7 ("strongly agree"). Higher scores indicate higher employee perceptions of leader empowering behaviors. All Cronbach's alpha reliability coefficients for scores on the six-factor model are acceptable (range = .82 to .90). All standardized factor coefficients are greater than .78 with the exception of Item 6 (.65) and Item 12 (.62). There is moderate variability in the scales as indicated by the standard deviations (SDs = 0.99 to 1.37). The inter-factor correlations range from .40 to .88. Overall, these results indicate that a six-factor model provides a good description of the relationships among the LEBQ items. For measurement of Agile leadership style, the LEBQ is a psychometrically sound instrument for the survey.

*Organization Structure, Customer Engagement and Team Autonomy*

The study employed three of the six dimensions of the Agile R&D units' organization (ARDO) questionnaire developed by Meier & Kock (2021).

These three dimensions are:

(1) The stakeholder integration (or customer integration) scale comprises four items assessing how the stakeholders are involved in the R&D unit's product development process.
(2) The team autonomy (or autonomy) scale reflects the extent to which the team can make its own decisions regarding how tasks should be done.
(3) The organization structure (or flat hierarchies) scale measures the organization structure using four items.

All items were measured on a ratio Likert scale with anchors at 1 ("does not apply at all") to 7 ("applies completely"). All values for Cronbach's alpha are above 0.70 suggesting high-scale reliability.

*Human Resource Practices*

The study used the opportunity-enhancing HRM practices of the HRM questionnaire developed by Zavyalova et al. (2018). These practices are referred to as Empowerment-enhancing HR practices by Gardner et al. (2011). The scale for opportunity-enhancing HRM practices includes four items. All items in the questionnaire take the form of "How often are the following HRM practices used in your organization?" with scales ranging from 1 (not used at all) to 5 (very often). Cronbach's alpha is 0.91, indicating good internal consistency of the construct.

## 6. ANALYSIS

The data collection for this research study was conducted via an online, anonymous survey in Qualtrics tool and shared it on the Prolific$^R$ data collection platform, various Scrum related forums including Scrum.org, subreddits on Agile Scrum and Programming, IEEE Software and the ASEM mailing list over a period of two weeks. The survey collected 137 responses and they were all complete since the survey tool forced a response to each survey question to proceed further. With four predictor variables, this is a sufficient sample size to allow useful analysis (Russo et al., 2021). The quantitative data generated was analyzed using multiple linear regression in R. The relationship between the independent variables – the human psychology factors pertaining to leadership style, organization structure, human resource practices, customer engagement and the dependent variable - Agile team autonomy is explained through multiple linear regression. As multiple items are linked to variables, the statistical analysis was performed using the median scores for each variable. One-way ANOVA and Pearson's correlation coefficient were used to demonstrate the existence (or nonexistence) of relationships between variables. Finally, an empirical model relating the four human psychology factor variables and the dependent variable of Agile team autonomy was constructed for the population.

## 7. RESULTS

Descriptive Statistics for the five study variables are presented in Exhibit 3.

Exhibit 3. Descriptive Statistics

| Variable | N | Mean | Standard Deviation | Variance | Median | Skewness | Kurtosis |
|---|---|---|---|---|---|---|---|
| Leadership Style (P1) | 137 | 5.730 | 1.010996 | 1.022112 | 6.0 | -1.15464 | 5.67752 |
| Stakeholder Collaboration (P2) | 137 | 4.650 | 1.483196 | 2.199871 | 5.0 | -0.26608 | 2.07652 |
| Organization Structure (P3) | 137 | 5.234 | 1.116473 | 1.246511 | 5.5 | -0.89903 | 4.53227 |
| Human Resource Practices (P4) | 137 | 5.102 | 1.431190 | 2.048304 | 5.0 | -0.85983 | 3.31575 |
| Team Autonomy (D) | 137 | 5.489 | 1.207311 | 1.457600 | 6.0 | -1.43366 | 5.59012 |

The skewness is negative, that indicates the distributions are left-skewed.

Since the kurtosis is mostly greater than 3, this indicates the distribution has more values in the tails compared to a normal distribution. The study constructs were treated as ordinal variables though the Likert scale was used. The ordinal values were assumed by the author to be an appropriate representation of the type of survey responses received.

Exhibit 4 shows a scatter plot of the variables in this study. Team Autonomy shows significant correlation with Leadership Style and Organization Structure. This is validated by the Pearson's correlation coefficients (r) summarized in Exhibit 5. In Social Sciences and Psychology research, r ≥ 0.30 is considered significant (Russo et al., 2021).

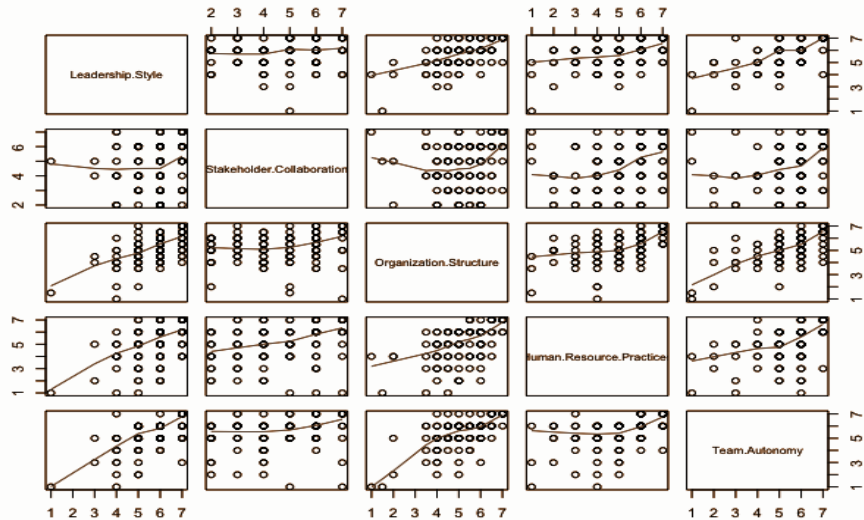

Exhibit 4. Scatter Plot

Exhibit 5. Pearson Correlation Coefficients

| Predictor Variable | Team Autonomy (D) | Cronbach's Alpha |
|---|---|---|
| Leadership Style (P1) | **0.39235398**** | 0.86 |
| Stakeholder Collaboration (P2) | 0.16260179 | 0.80 |
| Organization Structure (P3) | **0.34844601**** | 0.80 |
| Human Resource Practices (P4) | 0.09290828 | 0.91 |
| N | 137 | |

The p-value of the correlation between Team Autonomy with Leadership Style is 2.749650e-06 which is lower than α = 0.05. The p-value of the correlation between Team Autonomy with Organization Structure is 3.694010e-05 which is lower than α = 0.05. The low p-values suggest these relationships are representative of the population.

A linear model is thus constructed as follows:

*TeamAutonomy = $\beta_0$ + $\beta_1$\*LeadershipStyle + $\beta_2$\*StakeholderCollaboration + $\beta_3$\*OrganizationStructure + $\beta_4$\*HumanResourcePractices*

Based on multiple linear regression analysis, the linear model becomes:

*Team.Autonomy = 0.15626 + 0.45526\*LeadershipStyle +*
*0.09652\*StakeholderCollaboration + 0.36887\*OrganizationStructure +*
*0.06760\*HumanResourcePractices*

Given Adjusted R-squared = 0.5177 and close to Multiple R-squared = 0.5319, the model appears adequate for use. In Social Sciences and Psychology research, $R^2 \geq 0.50$ is considered acceptable (Russo et al., 2021).

A lack of fit test of this linear model was conducted using One-way ANOVA with a quadratic model and an interactions model. In both tests, the p-value was greater than α = 0.05 giving no reason to reject the null hypothesis. This implies there is no significant difference between the linear and quadratic model as well as between the linear and interactions model. Hence, the linear model appears to provide a reasonable fit to the data. A quadratic or interactions model is not justified.

## 8. HYPOTHESIS TESTING

*Hypothesis 1: No significant relationship exists between leadership style and team autonomy.*

This is measured by the LEBQ survey questionnaire, at a construct level, and the Agile Scrum Team's perception of the Leadership Style, at the construct level.

This hypothesis is rejected. There is a relationship between Leadership Style, at the construct level, and Team Autonomy, at the construct level ($r = 0.392$, p = 2.75e-06)
at α = 0.05, as shown in Exhibit 5.

*Hypothesis 2: No significant relationship exists between customer engagement and team autonomy.*

This is measured by the ARDO survey questionnaire, at a construct level, and the Agile Scrum Team's perception of Stakeholder Collaboration, at the construct level.
This hypothesis is accepted. There is no significant relationship between Stakeholder Collaboration, at the construct level, and Team Autonomy, at the construct level
($r = 0.163$, p = 6.05e-02) at α = 0.05, as shown in Exhibit 5.

*Hypothesis 3: No significant relationship exists between organization structure and team autonomy.*

This is measured by the ARDO survey questionnaire, at a construct level, and the Agile Scrum Team's perception of the Organization Structure, at the construct level.
This hypothesis is rejected. There is a relationship between Organization Structure, at the construct level, and Team Autonomy, at the construct level ($r = 0.348$, p = 3.69e-05) at α = 0.05, as shown in Exhibit 5.

*Hypothesis 4: No significant relationship exists between human resource practices and team autonomy.*

This is measured by the HRM survey questionnaire, at a construct level, and the Agile Scrum Team's perception of the Human Resource Practices, at the construct level.
This hypothesis is accepted. There is no significant relationship between Human Resource Practices, at the construct level, and Team Autonomy, at the construct level ($r = 0.093$, p = 2.86e-01) at α = 0.05, as shown in Exhibit 5.

Construct validity was established through Confirmatory Factor Analysis (CFA) in R. Factor loading values of at least 0.4 are considered adequate for this research and may be used to measure construct validity (MacCallum et al., 1999). LEBQ factor loading values, shown in Table 8, are greater than 0.4 for each question which is a confirmation of construct validity (MacCallum et al., 1999).

The first hypothesis seeks to determine whether there is a correlation between leadership style and team autonomy. Based on data from all 137 responses collected, it is evident that there is a correlation between the two variables. The leadership questions emphasize distributed leadership style. It can be inferred that agile teams regard distributed leadership to be a critical enabler for attaining team autonomy.

The third hypothesis seeks to determine whether there is a correlation between organization structure and team autonomy. Based on data from all 137 responses collected, it is evident that there is a correlation between the two variables. The organization structure questions enquire about the extent of hierarchy and speed of decision making in the organization. It can be inferred that team autonomy is adversely impacted by deep management hierarchy and power distance.

The second and fourth hypotheses were not rejected since there was no significant influence of customer engagement or human resource practices on team autonomy. Customer awareness and adoption of Agile Scrum methodology is still low in organizations and is a greenfield area for growth of Agile Scrum. Most Human Resource departments are still learning how to incentivize team practices and deemphasize individual performance. This remains an uncharted territory for Agile Scrum practitioners.

## 9. DISCUSSION AND IMPLICATIONS

Majority of Agile project management research has focused on the agile scrum process and tools but very little focus on the psychological factors that influence emergence of agile team autonomy. We have seen a lot of agile methodology but have very little understanding of the psychological factors that contribute to the emergence of autonomy in agile teams. Adoption of agile processes and tools alone is not enough to bring in agile team autonomy. We must develop a holistic understanding of the sources of psychological antecedents to agile team empowerment. It is therefore not a surprise that bureaucratic companies seem to struggle in their agile transformation (Moe et al., 2009; Nerur et al., 2005). Distributed leadership behavior is found to be a key success factor for fostering an agile self-organized team (Gren et al., 2019).

This research study comes at an opportune time and confirms that distributed leadership can indeed neutralize the adverse effect of uncooperative stakeholders, hierarchical organization structure and outdated human resource practices on agile scrum team autonomy. Software organizations can start to experimentally ascertain that adopting these recommendations will increase team autonomy and eventually agile project success. This study has provided useful insights into human psychology factors that are under the organization iceberg but still impede team autonomy. This will help organizations to design the right incentive programs to mitigate risks to team agility in software firms.

The study behooves leaders in the organization to make a sincere effort to empower Agile scrum teams for making their daily decisions with minimum guidance. Such leadership behavior will foster innovation, problem-solving agility and resilience in Agile Scrum teams and ultimately lead to a resilient organization that can weather any market storm. Additionally, if teams are finding creative ways to solve problems, new opportunities and increased organizational efficiencies may arise. This may remove the need to have deep hierarchies to achieve worker efficiency. Leadership can devote more time to surveying the future and understanding the business environment. The software engineering domain has become highly complex and the ability to create pathbreaking products needs complex, unbridled human thought. As the work of software engineers becomes more complex, more decisions must be made at the team level than at the leadership level. It is crucial to empower Agile Scrum teams to make independent judgment calls so that software firms can deploy new technologies and effectively manage existing technologies. Acting and offering Agile Scrum teams the opportunity to grow will evolve their role, strengthen their loyalty towards the organization and promote employee retention.

## 10. LIMITATIONS AND FUTURE RESEARCH

It is hard to establish strong correlations in research studies in the fields of social sciences and organizational psychology. This research likewise did not establish cause and effect. The correlations cited demonstrated a relationship. Further research may develop a more robust understanding of the relationships between the four human psychology factors and Agile Scrum team autonomy. The study shows how the human psychology factors resulting from distributed leadership and organization hierarchy can influence Agile Scrum team autonomy. It is also evident from the $R^2 = 0.52$ value that there are more human psychology factors in play in an organization like culture and organization history that may have an influence on Agile Scrum team autonomy.

This research was limited to software engineers with Agile experience in the United States. It is likely that a specific set of biases based on the knowledge, training and experience of a U.S. based software engineer may be present. While this research provides insight into the influence of four human psychology factors on the autonomy of Agile Scrum teams, it is by no means an exhaustive coverage of all factors. There may additionally be an opportunity to revisit the effect of these four factors on team autonomy in a remote vs. onsite workplace setting. The study did not ask for personal information from respondents like age, gender, extroversion, level of education and overall years of work experience. There may be hidden variables in this context that are waiting to be revealed. There is a risk of the data and findings becoming less relevant after a period especially with re-organizations and layoffs. Organization history may be a hidden variable but was not in scope for this research study.

This research elicited responses related to an agile scrum team member's perception about leadership style, stakeholder collaboration, organization structure and human resource practices as enablers for team autonomy. There could be other perspectives to consider here like that of the people manager, project manager, scrum master, product owner and customer that may prove to be insightful. Another, more complex approach, would involve the development of a longitudinal study to determine the long-term influence of the four human psychology factors on agile scrum team autonomy and project success. This study was focused on the influence of four specific human psychology factors on Agile Scrum team autonomy. Future studies can explore the existence and impact of additional human psychology factors on Agile Scrum team autonomy in the U.S. software industry. This research study can be extended to other countries with large numbers of software engineers like Romania, Brazil, India, and Singapore. This will enable the exploration of a country's culture as a human psychological factor affecting team autonomy. The cultural dimensions of power distance and individualism can vary with country and may have an influence on the emergence of team autonomy.

## 11. CONCLUSION

This research study comes at an opportune time and confirms that supportive and collaborative leadership can indeed neutralize the adverse effect of uncooperative stakeholders, hierarchical organization structure and outdated human resource practices on Agile Scrum team autonomy. Software organizations can start to experimentally ascertain that adopting these recommendations will increase team autonomy and eventually Agile project success. This study has provided useful insights into human psychology factors that are under the organization iceberg but still impede team effectiveness. This will help organizations to design the right incentive programs to mitigate risks to organization agility in software firms. This research addresses hitherto neglected topics in Agile teams' research. The findings will add value to the Engineering Management domains of Project Management, Leadership & Organizational Management and Management of Technology, Research, and Development.